\documentclass[preprint,nofootinbib,12pt]{revtex4-1}
\pdfoutput=1

\usepackage[utf8]{inputenc}

\usepackage{fullpage}

\usepackage{hyperref}

\usepackage{amsmath}
\usepackage{amssymb}
\usepackage{latexsym}
\usepackage{bbold}
\usepackage{color}

\usepackage{boxedminipage}

\usepackage{ifpdf}

\usepackage{slashed}

\usepackage{tikz}
\usetikzlibrary{positioning}
\usetikzlibrary{decorations.pathmorphing}
\usetikzlibrary{decorations.markings}
\usetikzlibrary{snakes}

\newcommand{\GeV}{{\rm GeV}}

\newcommand{\TeV}{{\rm TeV}}

\newcommand{\beq}{\begin{equation}}
\newcommand{\eeq}{\end{equation}}
\newcommand{\beqa}{\begin{eqnarray}}
\newcommand{\eeqa}{\end{eqnarray}}

\newcommand{\lsim}{\mathrel{\rlap{\lower4pt\hbox{\hskip1pt$\sim$}}
    \raise1pt\hbox{$<$}}}         
\newcommand{\gsim}{\mathrel{\rlap{\lower4pt\hbox{\hskip1pt$\sim$}}
    \raise1pt\hbox{$>$}}}         

\definecolor{myblue}{rgb}{0.2,0.2,0.7}
\definecolor{myred}{rgb}{0.7,0.2,0.2}

\begin{document}

\tikzset{ 
  scalar/.style={dashed},
  scalar-ch/.style={dashed,postaction={decorate},decoration={markings,mark=at
      position .5 with {\arrow{>}}}},
  fermion/.style={postaction={decorate}, decoration={markings,mark=at
      position .5 with {\arrow{>}}}},
  gauge/.style={decorate, decoration={snake,segment length=0.2cm}},
  gauge-na/.style={decorate, decoration={coil,amplitude=4pt, segment
      length=5pt}}
}


\vspace*{.0cm}

\vskip.5cm
\begin{center}
{\Large \bf Mesino Oscillation in MFV SUSY}
 
\end{center}
\vskip0.2cm

\begin{center}
{
{Joshua Berger}, {Csaba Cs\'aki}, {Yuval Grossman},
{\rm and}
{Ben Heidenreich}}
\end{center}

\begin{center}
{\it  Department of Physics, LEPP, 
Cornell University, Ithaca, NY 14853} \\
 
\vspace*{0.3cm}
{\tt  jb454,csaki,yg73,bjh77@cornell.edu}
\end{center}

\vglue 0.3truecm

\begin{center}
{\bf Abstract}
\end{center}
\vskip 3pt

$R$-parity violating supersymmetry in a Minimal Flavor Violation
paradigm can produce same-sign dilepton signals via direct sbottom-LSP pair
production.  Such signals arise when the sbottom hadronizes and the
resulting mesino oscillates into an anti-mesino.  The first bounds on
the sbottom mass are placed in this scenario using current LHC
results.

\section{Introduction}

The 2011 and 2012 data from the Large Hadron Collider (LHC) place severe
constraints on natural R-parity conserving models of supersymmetry
(SUSY) \cite{CMSSUSY:2012,ATLASSUSY:2012,CMSJETS:2012}.  While such models are not
excluded by the data, if they are to solve the hierarchy problem of
the Standard Model (SM), they are forced to have either non-generic
spectra where only third-generation squarks are light~\cite{Essig:2011qg,Kats:2011qh,Brust:2011tb,Papucci:2011wy,StopSeiberg}
or nearly degenerate particles, either in the form of stealth SUSY~\cite{Fan:2011yu}
or a squashed~\cite{LeCompte:2011fh} spectrum.
On the other hand, the stubborn agreement between SM
predictions and observations in channels with large missing transverse
energy (MET) cuts may indicate that the assumption of exact 
R-parity conservation is incorrect. 

Models with R-parity violation (RPV) have been proposed since the
early days of SUSY~\cite{RPV}. More recently a possible connection
between the problems of baryon and lepton number violation and large
flavor changing operators was
highlighted~\cite{NikolidakisSmith,Csaki:2011ge}.  The assumption of
Minimal Flavor Violation (MFV) has been shown to be sufficient to
prevent both rapid nuclear decay (and other baryon-number violating
processes) and large corrections to flavor observables in the $B$,
$D$, and $K$ systems.  In models of MFV SUSY, sparticles are pair
produced as in R-parity conserving models, while the lightest
supersymmetric partner (LSP) is generally unstable on collider scales
and will decay via the baryon-number violating $u^c d^c d^c$
superpotential term.  In this paper, we investigate one of the
interesting scenarios that can arise in the model
of~\cite{Csaki:2011ge}.

In MFV SUSY models it is particularly compelling to consider the case
when the LSP is a third generation squark: naturalness requires light
third generation squarks in general, and we will see below that in the
MFV scenario there is a high probability for this to be actually
realized, due to the large top Yukawa coupling.  The phenomenology of
this scenario is also particularly rich, as the lifetime of an LSP
stop or sbottom is long enough that the squark hadronizes to form a
mesino by binding to a light quark pulled from the vacuum.  It is,
however, usually sufficiently short-lived to decay before reaching the
detector. Observing squark production is challenging in this
scenario, due to the lack of any obvious handles on the events, such
as missing energy or displaced vertices. Instead we will make use of
the idea of mesino-antimesino oscillations, following Sarid and Thomas~\cite{Sarid:1999zx}.
 We will demonstrate that sbottom-LSP
pair production often allows for mesino-antimesino oscillations,
which may lead to same-sign dilepton signals.

A sbottom LSP decays dominantly to a top quark and a strange
quark~\cite{Csaki:2011ge}.  If one of the sbottom mesinos oscillates
before decaying, the tops will be of the same charge, and if both tops
decay leptonically this leads to same-sign leptons.  These events
would also contain $b$ quarks from the top decays, providing further
handles on the event. Recently, CMS searched for such events and
placed bounds on their cross sections~\cite{CMS:2012}. We will show
that this CMS search already places some bounds on sbottoms, which
should improve significantly with more data.

The rest of this paper is organized as follows.  In Section 2, we
study the typical squark spectra in MFV SUSY scenarios and demonstrate
that the stop and sbottom are most often the lightest squarks.  In
Section 3, we present a calculation of the decay rate and oscillation
time for a squark LSP in MFV SUSY and show that a significant
oscillation probability is possible and occurs frequently.  In Section 4, we
comment on the sensitivity of existing LHC searches to this
scenario. We conclude in Section 5. The details of the calculation of the mesino-antimesino
oscillation rate is given in the appendix.

\section{MFV Squark Spectra}
\label{sec:mfv-squark-spectra}

In an MFV SUSY model the LSP decays and we are not restricted to models
with a neutralino LSP, whereas the phenomenology of the model will depend on the identity
of the LSP.  In particular, the LSP can be colored and, as we consider
below, can be a squark.

MFV requires that all flavor violation be proportional to the
appropriate combination of Yukawa matrices, which are treated as
spurions of the flavor symmetry.   The squark mass 
matrices are then required to have the following form~\cite{Csaki:2011ge}:
\begin{equation}
  \label{eq:up-squark-mass}
  M_{\tilde{u}}^2 = \begin{pmatrix}
    m_{\tilde{q}}^2 \mathbb{1} + (a_q+v_u^2) Y_u Y_u^\dagger + b_q Y_d
    Y_d^\dagger + D_{u_L} & A_u Y_u \\ 
    A_u^* Y_u^\dagger & m_{\tilde{u}}^2 \mathbb{1} + (a_u+v_u^2) Y_u^\dagger
    Y_u + D_{u_R}
  \end{pmatrix}\, ,
\end{equation}
and
\begin{equation}
  \label{eq:down-squark-mass}
  M_{\tilde{d}}^2 = \begin{pmatrix}
    m_{\tilde{q}}^2 \mathbb{1} + a_q Y_u Y_u^\dagger + (b_q+v_d^2) Y_d
    Y_d^\dagger + D_{d_L} & A_d Y_d \\ 
    A_d^* Y_d^\dagger & m_{\tilde{d}}^2 \mathbb{1} + (a_d+v_d^2) Y_d^\dagger
    Y_d + D_{d_R}
  \end{pmatrix}\,.
\end{equation}
The $D$ terms are automatically flavor diagonal and given by   
\begin{equation}
  \label{eq:d-term-squark}
  D_L = \left(T^3 - Q s_w^2\right) \cos(2\beta)\, m_Z^2 \,, \qquad  D_R = Q s_w^2 \cos(2\beta)\, m_Z^2 \,,
\end{equation}
where $Q=+2/3$ ($-1/3$) and $T^3=+1/2$ ($-1/2$) for the up-type (down-type) squarks, $s_w$ is the sine of the Weinberg angle, $\tan \beta$ is the ratio of the Higgs VEVs, and $m_Z$ is the mass of the $Z$.  The parameters $m_i^2$, $a_i$, $b_i$, and $A_i$ arise from supersymmetry breaking, and we
therefore expect them to be of order 
$m_{\text{soft}}^2$. 

\begin{figure}[tb]
  \centering
  \includegraphics[height=0.25\textheight]{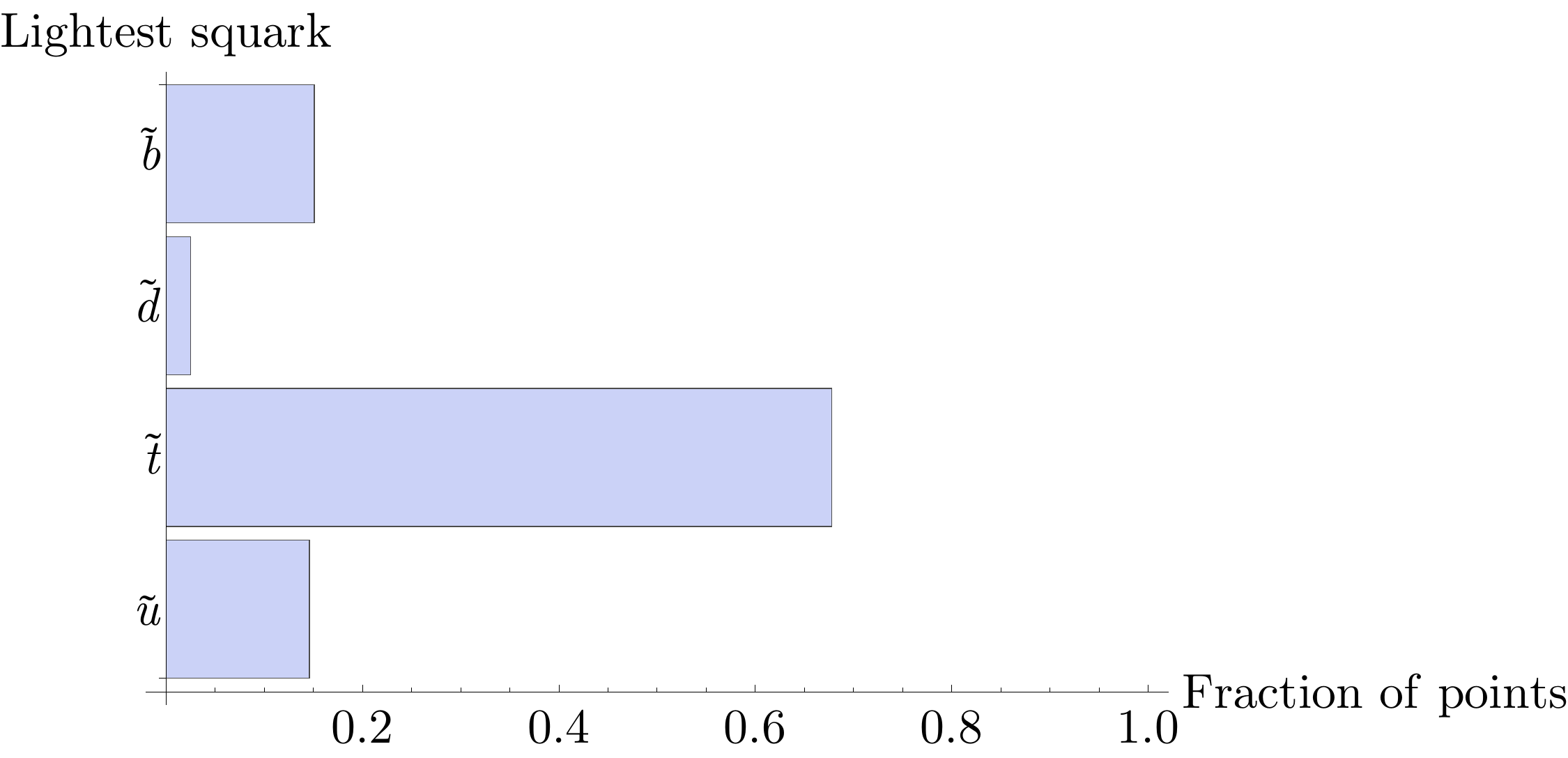}
  \caption{Distribution of lightest squark flavor over a random
sampling of MFV SUSY parameter space.}
  \label{fig:flav-distrib}
\end{figure}
Given these constraints, we can perform a scan over the parameter
space that determines the squark spectrum.  We select random values
for the undetermined dimension-two parameters uniformly in
$[-m_{\text{soft}}^2,m_{\text{soft}}^2]$.  For this scan, we choose
$m_{\text{soft}} = 1~\TeV$ and $\tan\beta = 10$. The overall result is
not very sensitive to this choice. We impose the constraint that the
smallest eigenvalues of both squark mass matrices be greater than the
top mass, $m_t \approx 175~\GeV$.  In general, left-right mixing is not
too large and we therefore use notation where $\tilde b_L$ refers to the
mass eigenstate of the sbottom that is mostly a left-handed
sbottom. We also impose that the lightest stop-like squark be lighter
than $500~\GeV$ as demanded by naturalness.  Under these conditions,
the distribution of lightest squark flavors is given by
Fig.~\ref{fig:flav-distrib}.

We observe
that roughly $85\,\%$ of parameter points have a third-generation
lightest squark, out of which $15\,\%$ have a sbottom squark at the
bottom of the spectrum.  The large likelihood of a third generation
lightest squark can be explained by the relatively large left-right
mixing for this generation.  This mixing tends to drive the mass of
the lighter third generation squark down, making it more likely to be
lightest overall.  (There is also a significant contribution from the
naturalness cut, since requiring one light stop tends to reduce the
incidence of both stops being made heavy by a large positive $a_q$.)
Note that at large $\tan\beta$ this effect is enhanced for the
sbottoms, making it even more likely to get a sbottom LSP.  It is
therefore natural to consider signatures of a sbottom LSP in MFV SUSY
and we do so from this point on.

\section{Mesino oscillation in MFV SUSY}
\label{sec:mesino-oscill-mfv}

The MFV SUSY scenario offers a rich phenomenology due to the naturally
small decay width of the LSP, a consequence of approximate $R$-parity
conservation.  The couplings are sufficiently small to yield LSP
lifetimes that are longer than the timescales of SM short-distance
physics, such as hadronization, yet often shorter than the timescales
set by macroscopic distances in the LHC detectors.  In this
intermediate  range, it can be difficult to construct observables that
are not overwhelmed by SM background.  If the LSP carries color,
however, then it lives sufficiently long to hadronize, an intriguing
possibility.  This process can yield additional phenomena that allow
for efficient selection of SUSY events. 

The case of a sbottom LSP is particularly fruitful.  If the gluino is
heavy, the dominant SUSY production mode will be sbottom pair
production.  The dominant decay of the sbottom in MFV SUSY is to top
and strange.  The top has a leptonic decay mode, which already
suppresses many SM backgrounds.  As we show, the fact that the sbottom
hardronizes allows for the possibility of sbottom oscillations, which
lead, some fraction of the time, to same-sign lepton events.

While other squark flavors can also oscillate, this turns out to be
parametrically rarer.  In addition, up-type squark LSPs do not decay
leptonically, precluding the possibility of a same-sign dilepton
signature. We do not consider these other possibilities further in
this work.

We also do not consider the case where gluino pair production is significant. This would lead to additional same-sign lepton events due
to the Majorana nature of the gluino, providing a background to the case we are considering.

\begin{figure}[tb]
  \centering
  \begin{tikzpicture}
    \draw[scalar-ch] (0,0) node[left] {$\tilde{b}_1$} -- (2,0);
    \draw[fermion] (3.414,1.414) node[right] {$t^c$} -- (2,0);
    \draw[fermion] (3.414,-1.414) node[right] {$s^c$} -- (2,0);
 \end{tikzpicture}
  \caption{The leading diagram for the $R$-parity violating sbottom decay.}
  \label{fig:squark-dec}
\end{figure}
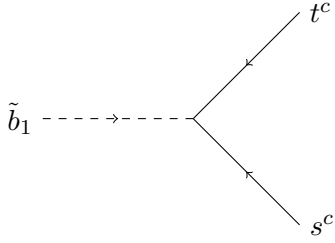
We begin by calculating the sbottom decay width. We denote the
lightest sbottom mass eigenstate by $\tilde b$. Its decay width
depends on an overall (generically order 1) coefficient that we denote
by $\lambda^{\prime\prime}$.  The Lagrangian terms that gives the decay
is~\cite{Csaki:2011ge}:
\begin{equation}
  \label{eq:sbottom-decay-term}
  \mathcal{L} = - (\lambda^{\prime\prime})^*  \epsilon_{ijk}
  \frac{m_{q}}{v c_\beta} U_{q_R,1}^D \frac{m_{u,j}}{v s_\beta}
  V_{j^\prime j} \frac{m_{d,k}}{v c_\beta} \tilde{b}_1 u^{c\dagger}_{j^\prime} d^{c\dagger}_{k} \sim -
  (\lambda^{\prime\prime})^* V_{td} \frac{m_b m_s m_t}{v^3 c_\beta^2
    s_\beta} \tilde{b}_1 t^{c\dagger} s^{c\dagger},
\end{equation}
where $v = 174~\GeV$, $V$ is the CKM matrix, and we use $\tilde{b}_1$ to
denote the lightest down-type squark, which we assume is
predominantly sbottom-like.  The mixing matrix
$U^D$ is defined such that
\begin{equation}
  \label{eq:mixing-mat-def}
  \tilde{q}_q = U^D_{q,i} \tilde{q}_i.
\end{equation}
In this notation, $\tilde{q}_q$ are the squark flavor-basis fields in the
mass basis of the quarks and $\tilde{q}_i$ are the squark mass-basis
fields.  The approximation is valid if the lightest
sbottom is mostly right-handed.  Otherwise, there is an additional
suppression from the left-right mixing.  The partial decay width can
then be calculated using the diagram in Fig.~\ref{fig:squark-dec}.
The result (neglecting the mass of the down quark in the phase space
integral) is:
\begin{equation}
  \label{eq:sbottom-width-calc}
  \Gamma = \sum_{j^\prime, k} \frac{1}{32\pi^2} \left| (\lambda^{\prime\prime})^* \sum_{i,j,q} \epsilon_{ijk}
  \frac{m_q}{v c_\beta} U_{q_R,1}^D \frac{m_{u,j}}{v s_\beta}
  V_{j^\prime j} \frac{m_{d,k}}{v c_\beta}\right|^2 m_{\tilde{b}} \left(1 - \frac{m_{u,j^\prime}^2}{m_{\tilde{b}}^2}\right)^2.
\end{equation}

To gain some intuition about sbottom LSP decays, we now make a few
approximations.  The decay is dominated by $\tilde{b}\to
t^c s^c$ provided there is sufficient phase space.
In the interesting segment of parameter space, the LSP is made up
almost entirely of some admixture of the left-handed and right-handed
sbottom, so that the decay width is approximately:
\begin{eqnarray}
  \label{eq:sbottom-mixing-factor}
  \Gamma  & \approx &
  \frac{1}{32\pi^2} |\lambda^{\prime\prime}|^2 \sin^2\theta \frac{m_b^2 m_s^2
  m_t^2}{v^6 c_\beta^4 s_\beta^2} |V_{td}|^2 m_{\tilde{b}} \left(1 -
  \frac{m_t^2}{m_{\tilde{b}}^2}\right)^2 \nonumber\\ ~ & \sim & (2.6
  \times 10^{-10}~\GeV) |\lambda^{\prime\prime}|^2 \sin^2\theta
  \left(\frac{t_\beta}{10}\right)^4
  \left(\frac{m_{\tilde{b}}}{300~{\rm GeV}}\right),
\end{eqnarray}
where $\theta$ is the left-right mixing squark mixing angle.

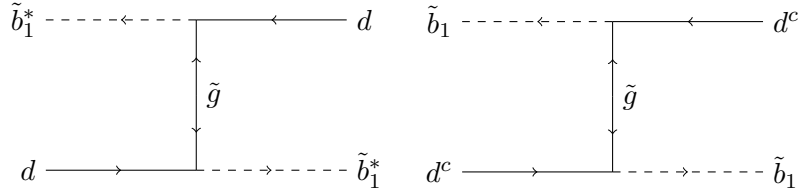
\begin{figure}[tb]
  \centering
  \begin{tikzpicture}
    \draw[scalar-ch] (2,0) -- (0,0) node[left] {$\tilde{b}_1^*$};
    \draw[fermion]  (4,0) node[right] {$d$} -- (2,0);
    \draw[scalar-ch] (2,-2) -- (4,-2) node[right] {$\tilde{b}_1^*$};
    \draw[fermion] (0,-2) node[left] {$d$} -- (2,-2);
    \draw[fermion] (2,-1) node[right] {$\tilde{g}$} -- (2,0);
    \draw[fermion] (2,-1) -- (2,-2);
  \end{tikzpicture}\hspace{0.2cm}
\begin{tikzpicture}
    \draw[scalar-ch] (2,0) -- (0,0) node[left] {$\tilde{b}_1$};
    \draw[fermion]  (4,0) node[right] {$d^c$} -- (2,0);
    \draw[scalar-ch] (2,-2) -- (4,-2) node[right] {$\tilde{b}_1$};
    \draw[fermion] (0,-2) node[left] {$d^c$} -- (2,-2);
    \draw[fermion] (2,-1) node[right] {$\tilde{g}$} -- (2,0);
    \draw[fermion] (2,-1) -- (2,-2);
  \end{tikzpicture}  
  \caption{Diagrams for sbottom mesino oscillation mediated by a gluino. There are also similar diagrams mediated by neutralino exchange, which can become important if the gluino mass is very large.}
  \label{fig:mesino-osc}
\end{figure}
The sbottom decay rate is much less than the hadronization
scale $\Lambda_{\text{QCD}} \sim 0.2~\GeV$. Thus, the sbottom squark
will hadronize before decaying to form fermionic mesino bound states
$\tilde{B}_q =\tilde{b}^* q$ and $\tilde{B}^c = \tilde{b}
q^c$.  If $q = d,s$, then the mesino is neutral, opening up the
possibility for mesino oscillations, first discussed
in~\cite{Sarid:1999zx}.  Since few details of the calculation of the oscillation rate were
given in
\cite{Sarid:1999zx}, we elaborate on it in
Appendix~\ref{sec:appendix}, explaining the necessary approximations. Our final result, eq.~(\ref{eq:deltam-mesino}), is in broad agreement with that of~\cite{Sarid:1999zx}, and we restate it here:
\begin{equation}
  \label{eq:mesino-osc}
 \Delta m= \omega = g_s^2 \left|(U^{D}_{d_L,1})^2 + (U^{D}_{d_R,1})^2\right| f_{\tilde{B}}^2 \left(1 - \frac{1}{N_c^2}\right)
  \frac{m_{\tilde{g}}}{m_{\tilde{g}}^2 - m_{\tilde{b}}^2}.
\end{equation}

This result depends on the nature of the spectator quark.  
We can use MFV to approximate the ratio of the oscillation rates as we have
$|U_M^{q1}|\propto |V_{tq}V_{tb}|$ for $M=L,R$.  In this
approximation, we get:
\begin{equation}
\label{eq:ckm-ratio}
\frac{\omega_s}{\omega_d} \approx \left|\frac{V_{ts}}{V_{td}}\right|^2 \approx 23
\end{equation}
The dependence of this ratio on the dimension-two parameters of the squark mass matrix is generically very weak.

With this factor in mind, we consider oscillation of the sbottom-down
mesino.  The oscillation rate can be estimated by
\begin{eqnarray}
  \label{eq:mesino-osc-approx}
  \omega & \approx & \frac{f_{\tilde{B}}^2}{2} \cos^2\theta\, |V_{td} V_{tb}^*|^2
  \frac{m_t^4}{v^4 s_\beta^4} \frac{m_{\tilde{g}}}{m_{\tilde{g}}^2  - m_{\tilde{b}_1}^2} \nonumber\\
  ~ & \sim & (4 \times 10^{-12}~\GeV)
  \left(\frac{f_{\tilde{B}}}{28.7~{\rm MeV}}\right)^2 \cos^2\theta\, \left(\frac{1000~\GeV}{m_{\tilde{g}}}\right).
\end{eqnarray}
These results are not too far from the decay rates,
eqs.~\eqref{eq:sbottom-mixing-factor}, but with different parametric
dependence. Thus, we expect some parts of parameter space
where the oscillation rate is comparable to or larger than the decay
rate, leading to appreciable mesino oscillations. 

\begin{figure}[tb]
  \centering
  \includegraphics[height=0.25\textheight]{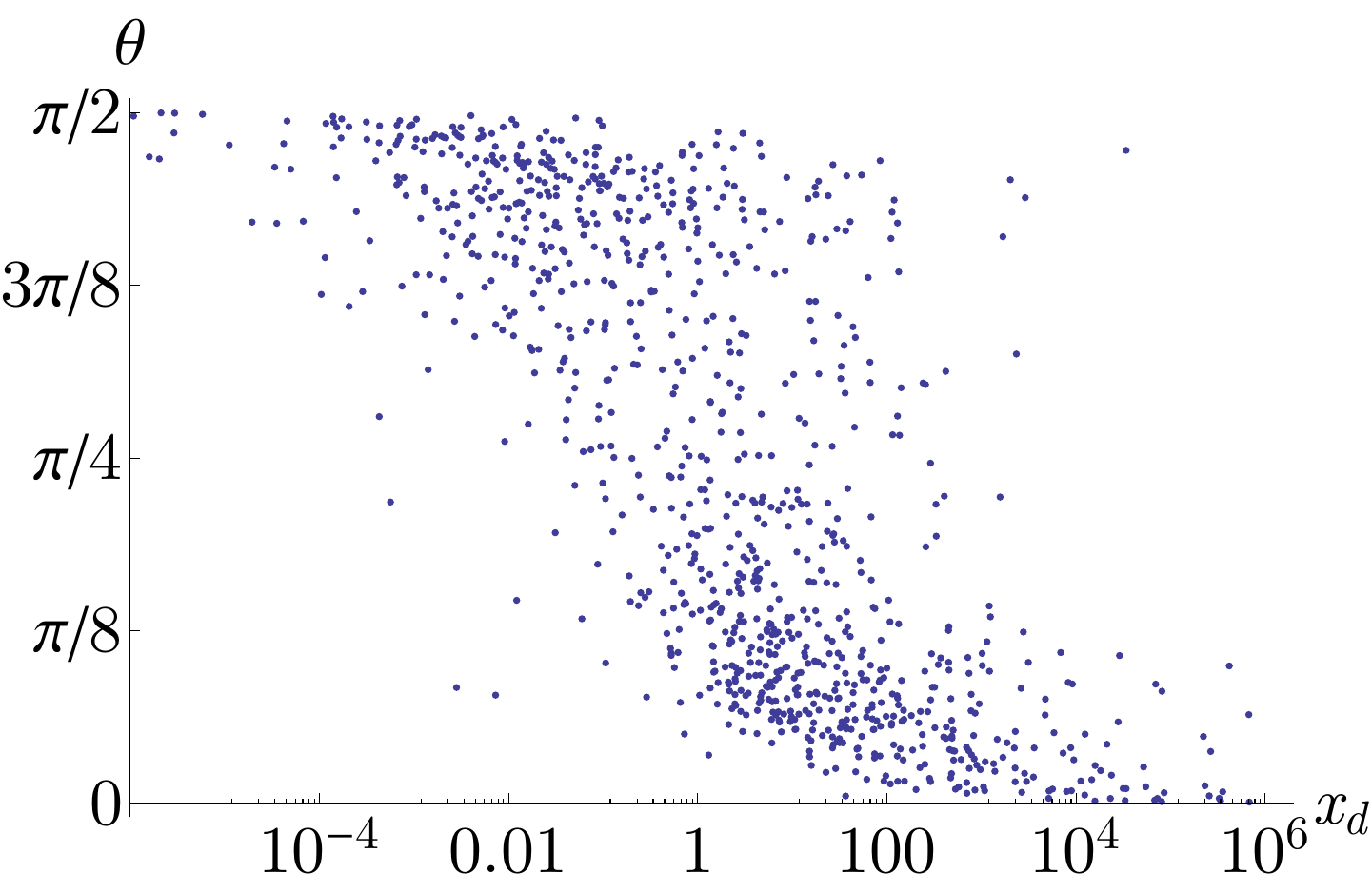}
  \caption{Oscillation parameter $x_d = \Delta m_{\tilde{B}_d}/\Gamma$ and left-right
    mixing angle $\theta$ resulting from a scan over parameter space, where $\theta = 0$ corresponds to a pure left-handed LSP.}
  \label{fig:osc-param-scan}
\end{figure}
To get a better sense of how common such a phenomenon is, we define
the oscillation parameter
\begin{equation}
  \label{eq:osc-param}
  x \equiv \frac{\Delta m}{\Gamma}.
\end{equation}
The time-integrated probability for a sbottom mesino to oscillate into an anti-sbottom mesino before decaying is
\begin{equation}
  \label{eq:osc-prob}
  p(x) \equiv P(\tilde{B} \to \tilde{B}^c) = \frac{x^2}{2(1 + x^2)}.
\end{equation}
The oscillation probability is small for $x\ll 1$ and becomes appreciable near $x \sim 1$, whereas for $x \gg 1$ the
$\tilde{B}$ oscillates very rapidly, and the mesino contains an equal mixture of sbottom and anti-sbottom components. We scan over parameter space using the same procedure as in Section
\ref{sec:mfv-squark-spectra}, selecting points with a
sbottom LSP and calculating $x_d$ (the $\tilde{B}_d$ oscillation parameter) and $\theta$ for each such point.
The results of the scan are shown in Fig.~\ref{fig:osc-param-scan}. We observe that $x_d > 1$ in a significant
portion of parameter space, particulary when the LSP is
predominantly left-handed.

\begin{figure}[tb]
  \centering
  \begin{tikzpicture}
    \draw[scalar-ch] (0,0) node[left] {$\tilde{b}_1$} -- (1,0);
    \draw[fermion] (1,0) -- node[above] {$\tilde{B}_1$} (2,0);
    \draw[fermion] (3,0) -- node[above] {$\tilde{B}_1^c$} (2,0);
    \draw[fermion] (3,0) --  node[above left] {$t$} (4.414,1.414);
    \draw[fermion] (3,0) -- (5,0) node[right] {$s$};
    \draw[fermion] (4.414,1.414) -- (6.414,1.414) node[right] {$b$};
    \draw[gauge] (4.414,1.414) -- node[above left] {$W$}
    (5.828,2.828);
    \draw[fermion]  (7.242,4.242) node[above right]
    {$\ell^+$}  -- (5.828,2.828);
    \draw[fermion] (5.828,2.828) --  (7.828,2.828) node[right]
    {$\nu$};
    
    \draw[scalar-ch] (1,-1) -- (0,-1) node[left] {$\tilde{b}_1^*$};
    \draw[fermion] (3,-1) -- node[below] {$\tilde{B}_1^c$} (1,-1);
    \draw[fermion] (3,-1) --  node[below left] {$t$} (4.414,-2.414);
    \draw[fermion] (3,-1) -- (5,-1) node[right] {$s$};
    \draw[fermion] (4.414,-2.414) -- (6.414,-2.414) node[right] {$b$};
    \draw[gauge] (4.414,-2.414) -- node[below left] {$W$}
    (5.828,-3.828);
    \draw[fermion]  (7.242,-5.242) node[below right]
    {$\ell^+$}  -- (5.828,-3.828);
    \draw[fermion] (5.828,-3.828) --  (7.828,-3.828) node[right] {$\nu$};
 \end{tikzpicture}
  \caption{Diagram for $R$-parity violating sbottom decay that leads
to same sign leptons.}
  \label{fig:squark-chain}
\end{figure}
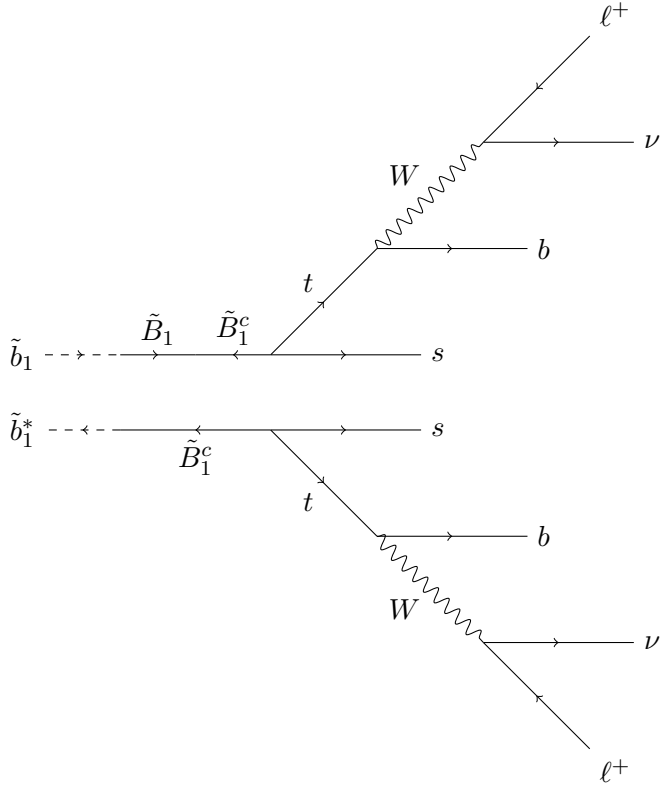
If the sbottom is the LSP and has a mesino oscillation time comparable
to or larger than its lifetime, then there is a very distinct
signature of direct sbottom pair-production.  The sbottoms will
hadronize and the resulting mesino may oscillate before decaying.  The mesino must be neutral for oscillations to occur, which
occurs when the spectator is a down or strange quark, or roughly half the time as estimated from the $B$ system.  If
exactly one of the mesinos oscillates before decaying, then the resulting
two halves of the final state will have the same charge.  Furthermore,
these final states each involve a top quark whose charge is easy to
tag if it undergoes a leptonic decay.  This  final state has
same-sign leptons, $b$ jets, and a small, but non-negligible, amount
of missing energy.  The entire chain is illustrated in
Fig.~\ref{fig:squark-chain}. The branching fraction for this mode is
given by
\begin{eqnarray}
  \label{eq:branching-golden} 
\text{Br}(\tilde{b}\tilde{b}^* \to  bb\ell^\pm \ell^\pm) &=&
\text{Br}(W \to \ell\nu)^2 f(x_d,x_s) \approx f(x_d,x_s)\times
 6.5\,\%, \nonumber \\
f(x_d,x_s) &=& 2 \sum_{i = d,s} h_i p(x_i) 
\left(1 - \sum_{j = d,s} h_j p(x_j) \right),
\end{eqnarray}
where $f(x_d,x_s)$ denotes the probability that exactly one of the two mesinos oscillates.
Here $h_i$ is the fraction of sbottoms that form mesinos with
spectator $i$ and we use the effective leptonic rate
for the $W$ which includes leptonic tau decays.  Note that
for $x_i\gg 1$ the rate is maximal, and since $h_d+h_s\approx 1/2$ we have $f(x_d,x_s)\approx 3/8$. Despite the modest
branching fraction, this decay mode will likely be the most sensitive
channel for discovering a sbottom LSP in MFV SUSY.

\section{Bounds from a CMS search}
\label{sec:bounds-from-cms}

CMS already has a search~\cite{CMS:2012} that is quite sensitive to
the above decay chain.  The same-sign dilepton and $b$ jets search
includes search regions with 0, $30~\rm{GeV}$ and $50~\rm{GeV}$ MET
cuts, all of which can be sensitive to our scenario due to the
neutrinos from leptonic top decays.  The relevant bounds from this
search are presented in Table~2 of~\cite{CMS:2012}.  In addition, they
present efficiency fits for the various cuts in terms of parton-level
objects, allowing for easy reinterpretation.  In this section, we use
this information to reinterpret their bounds in terms of MFV SUSY with
a sbottom LSP, and comment on future prospects.

\begin{figure}[tb]
  \centering \includegraphics[height=0.3\textheight]{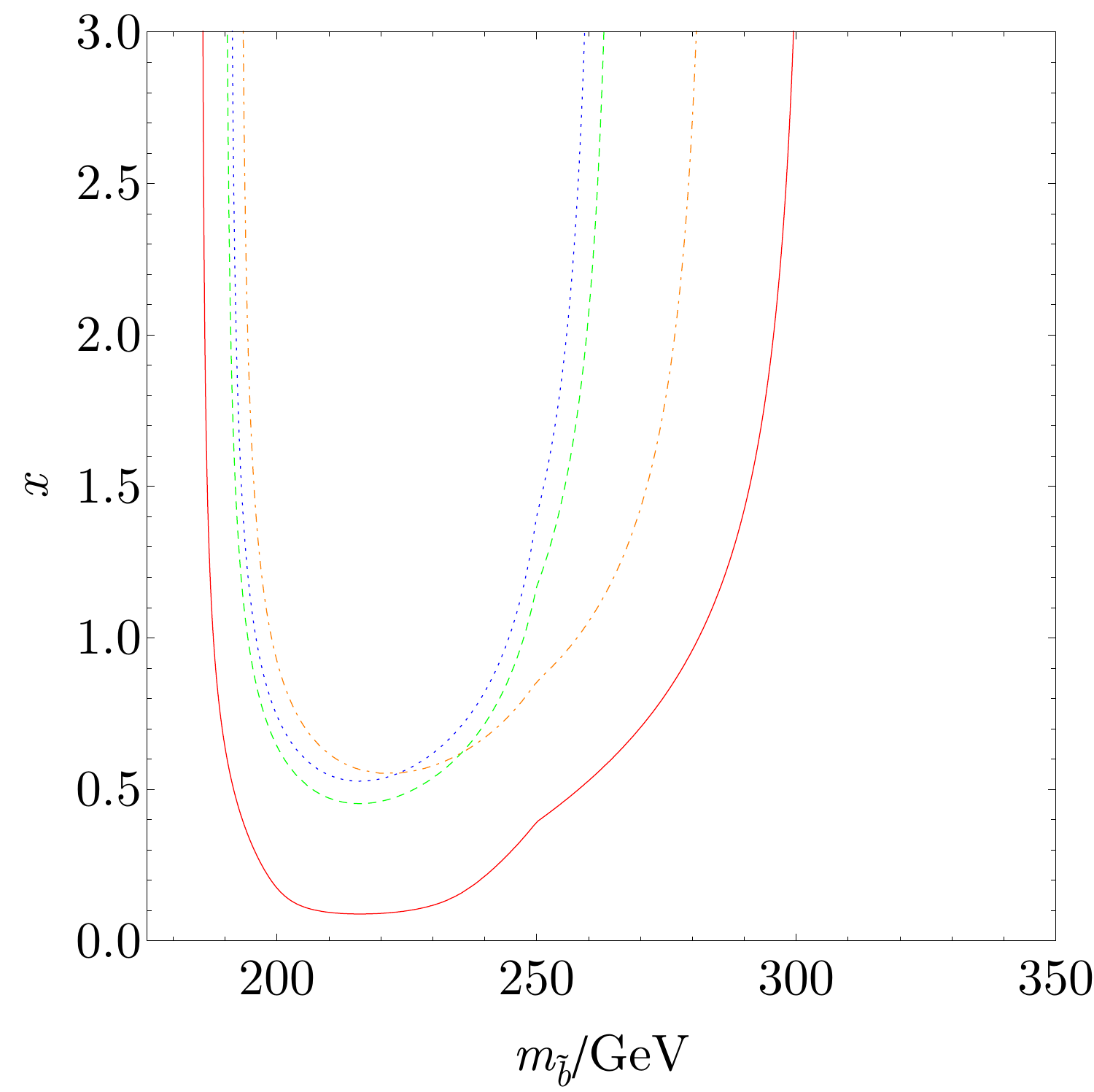}
  \caption{Recasted CMS bounds on sbottom direct production in terms
  of the sbottom mass and $x_d$.  Only the four most
  sensitive signal regions are shown: SR0 in dashed green, SR1 in
  dotted blue, SR2 in solid red, and SR4 in dash-dotted orange.  The
  most conservative upper limit on the number of new physics events is
  used for each search region, though varying this number has little effect on the bounds.}
  \label{fig:sbottom-bounds}
\end{figure}
To obtain a bound, we generated $pp \to \tilde{b} \tilde{b}^*$ at
$8~\TeV$ using Pythia 8 with all showering and hadronization turned
off.  The events are decayed at the parton-level.  The analysis cuts are
applied using the efficiencies presented in Section 6
of~\cite{CMS:2012}.  No mixing is introduced in event generation, but an
$x_d$-dependent factor is applied to the final efficiency to account for
the branching fraction to same-sign leptons.  The cross-section
for pair production is calculated at NLO using Prospino 2.1
\cite{Beenakker:1996ed,Plehn:2012}.  The resulting bounds are 
shown in Fig.~\ref{fig:sbottom-bounds}.  SR2, which counts only
positively chaged same sign pairs, yields the strongest
bound, since a presumed fluctuation in the data led to all observed
same-sign events having negatively charged leptons.  For a maximal
same-sign branching fraction, the exclusion extends between $180~\GeV$
and $305~\GeV$.

Note that obtaining same sign lepton events requires $x \gsim 1$. The
reason that there is sensitivity in the small $x_d$ region is due to
the possibility of producing strange mesinos. Even for $x_d \ll 1$, it
is possible to get $x_s>1$.

\section{Conclusions}
\label{sec:conclusions}

MFV SUSY is a compelling new paradigm for exploring supersymmetry
without $R$-parity that offers many new and challenging channels to
explore at the LHC.  A systematic study of the phenomenology of all
plausible scenarios in this framework is required to ensure full
sensitivity to weak-scale supersymmetry.  We have explored one
interesting scenario with a sbottom-like squark LSP.

Direct squark production will be essential for probing all possible
corners of natural SUSY parameter space.  Our work has demonstrated
that the LHC can be sensitive to directly produced sbottom LSPs in the
MFV SUSY scenario using the important fact that in this framework
strongly-interacting LSPs will hadronize. Using a CMS search for same
sign dileptons and $b$-jets, we have put a bound on sbottoms with
masses between 180 and 305 GeV that undergo large mesino oscillations,
which is a plausible scenario for the case with sbottom LSPs.

\begin{acknowledgments}
We thank Luke Winstrom for useful discussions, and Scott Thomas for correspondence on~\cite{Sarid:1999zx}. This research was
supported in part by the NSF grant PHY-0757868.
\end{acknowledgments}

\begin{appendix}
\section{Determination of the Mesino Oscillation Frequency}
\label{sec:appendix}

In this appendix, we present the details of the  calculation of the mesino oscillation
frequency, carefully listing all approximations as they enter.
Throughout the appendix, we will assume a down quark spectator, but
the results extend trivially to the strange quark case.  We further
denote the lightest squark by $\tilde{b}$ and assume that it is
sbottom-like.

In the quark and squark mass basis, there are two combinations of
sbottom and down quark that correspond to light mesino Weyl fermions:
\begin{equation}
  \label{eq:mesino-fields} 
\tilde{B}_1 \equiv \tilde{B} = \tilde{b}^* d\,,\qquad 
\tilde{B}_2 \equiv \tilde{B}^c = \tilde{b} d^c\,.
\end{equation}
The most general quadratic Lagrangian for these mesino fields is given
by:
\begin{equation}
  \label{eq:mass-terms-mesino}
  \mathcal{L} = \frac{1}{2} m_{ij} \tilde{B}^i \tilde{B}^j + {\rm h.c.}\,.
\end{equation}
Before including corrections
due to the gluino, the diagonal entries of $m_{i j}$ vanish, and the
two Weyl fermions combine to form a Dirac fermion.  The mass,
corresponding to the off-diagonal terms in
\eqref{eq:mass-terms-mesino}, is given to leading order by
\begin{equation}
  \label{eq:dirac-mass-leading}
  m_{12} = m_{\tilde{b}}\,.
\end{equation}
The leading corrections are of order $\Lambda_{\rm QCD}$, which we
neglect.

The diagonal elements $m_{11}$ and $m_{22}$, corresponding to Majorana masses for $\tilde{B}$ and $\tilde{B}^c$, are not in general equal, and are generated at leading order by tree-level gluino exchange, leading to an oscillation between mesinos and antimesinos. The oscillation frequency is equal to
the mass splitting between the two mass eigenstates, whose squared masses are the eigenvalues of $m^{\dagger} m$.
We take $m_{12}$ to be real by performing an appropriate field redefinition, in which case the eigenvalues of $m^\dagger m$ are given by 
\begin{equation}
  \label{eq:eigen-m2}
  \frac{1}{2} \left(|m_{11}|^2 + |m_{22}|^2 +2 m_{12}^2 \pm
    \sqrt{(|m_{11}|^2 + |m_{22}|^2 +2 m_{12}^2 )^2 - 4 |m_{12}^2 -
      m_{11} m_{22}|^2}\right)\,.
\end{equation}
To leading order in $m_{11}$ and $m_{22}$, the resulting mass
splitting is
\begin{equation}
  \label{eq:deltam-leading}
  \omega = \Delta m = |m_{11} + m_{22}^*|\,.
\end{equation}

We work at leading order in the heavy squark approximation.
Instead of determining $m_{11}$ and $m_{22}$ directly, we employ the simple and general formula:
\begin{equation}
\label{eq:gen-osc}
\omega = \frac{1}{m_{12}} |\langle \bar{\tilde{B}}(\vec{0},s) | \mathcal{H}_{\rm eff}(\vec{0}) | \tilde{B}(\vec{0},s) \rangle| \,,
\end{equation}
for $\omega \ll m_{12}$, where $\mathcal{H}_{\rm eff}(\vec{x})$ is the
effective Hamiltonian density generated by integrating out the gluino
and $| \tilde{B}(\vec{p},s) \rangle$ and $| \bar{\tilde{B}}(\vec{p},s)
\rangle$ denote one-particle mesino and antimesino states,
respectively, with momentum $\vec{p}$ and spin $s$ with no sum
over $s$. (We use the standard covariant normalization for
one-particle momentum eigenstates, $\langle \vec{p}|\vec{q} \rangle =
2 E_{\vec{p}} (2\pi)^3 \delta^{(3)} (\vec{p}-\vec{q})$.) The effective
Hamiltonian density from integrating out the gluino is:
\begin{equation} \label{eq:Heff}
\mathcal{H}_{\rm eff} =  \frac{C_L}{2} (\tilde{b}^{*} d) (\tilde{b}^{*} d) + \frac{C_R}{2} (\tilde{b} d^c) (\tilde{b} d^c)+h.c. \,,
\end{equation}
for coefficients $C_L$ and $C_R$ to be determined, where the color indices are contracted as indicated by the parentheses. Thus,
\begin{equation} \label{eq:omegabraket}
\omega = \frac{1}{m_{\tilde{b}}} \left|\frac{C_L}{2} \langle \bar{\tilde{B}} |(\tilde{b}^{*} d) (\tilde{b}^{*} d) | \tilde{B} \rangle+ \frac{C_R^{*}}{2} \langle \bar{\tilde{B}} |(\tilde{b}^{*} d^{c \dag}) (\tilde{b}^{*} d^{c \dag}) | \tilde{B} \rangle \right| \,.
\end{equation}
The structure is very similar to~\eqref{eq:deltam-leading}, and indeed the two terms within the absolute value in~\eqref{eq:omegabraket} are precisely $m_{\tilde{b}}$ times the Majorana masses which appear in~\eqref{eq:deltam-leading}.


To determine the $C_{L,R}$, we compare the short-distance amplitudes
for oscillation obtained using the MSSM Lagrangian and using the
effective Hamiltonian in \eqref{eq:Heff}.  The
MSSM gluino exchange amplitudes $\mathcal{M}_L$ and
$\mathcal{M}_R$ (Fig.~\ref{fig:mesino-osc}) are given by 
\begin{eqnarray}
  \label{eq:short-distance}
  \mathcal{M}_L & = & 
  2 [g_s^2 (U^{D*}_{d_L,1})^2] \left[\frac{m_{\tilde{g}} \delta^\alpha_\beta}{m_{\tilde{g}}^2 -
    m_{\tilde{b}_1}^2}\right] \left[t^a_{ij} t^a_{i^\prime j^\prime} +
  t^a_{ij^\prime} t^a_{i^\prime j}\right],\nonumber\\
\mathcal{M}_R & = &
  2 [g_s^2 (U^D_{d_R,1})^{2}] \left[\frac{m_{\tilde{g}} \delta^\alpha_\beta}{m_{\tilde{g}}^2 -
      m_{\tilde{b}_1}^2}\right] \left[t^a_{ij} t^a_{i^\prime j^\prime} + t^a_{ij^\prime} t^a_{i^\prime j}\right],
\end{eqnarray}
where we work in a basis where the gluino mass is real, and an overall factor of two arises since the gluino-quark-squark vertex
comes with a factor of $\sqrt{2}$.
The color factors in these amplitudes simplify to \cite{Terning:2006bq}:
\begin{equation}
t^a_{ij} t^a_{i^\prime j^\prime} + t^a_{ij^\prime} t^a_{i^\prime j} =
\frac{1}{2} \left(\delta_{ij^\prime} \delta_{i^\prime j} + \delta_{ij}
  \delta_{i^\prime j^\prime} - \frac{1}{N_c} \delta_{ij^\prime}
  \delta_{i^\prime j} - \frac{1}{N_c} \delta_{ij} \delta_{i^\prime
    j^\prime} \right) \,.
\end{equation}
The effective operators in~\eqref{eq:Heff} yield amplitudes:
\begin{equation}
  \label{eq:eff-calculation}
  \mathcal{M}_{L,R}^\prime = C_{L,R} (\delta_{ij^\prime} \delta_{i^\prime j} + \delta_{ij} \delta_{i^\prime j^\prime}) \,.
\end{equation}
By demanding that $\mathcal{M}_L$ ($\mathcal{M}_R$) from
\eqref{eq:short-distance} is equal to $\mathcal{M}_L^\prime$
($\mathcal{M}_R^\prime$) from \eqref{eq:eff-calculation}, we extract the coefficients $C_L$ and $C_R$:
\begin{equation}
  C_L = g_s^2 (U^{D*}_{d_L,1})^2
  \frac{m_{\tilde{g}}}{m_{\tilde{g}}^2 - m_{\tilde{b}_1}^2} \left(1 -
      \frac{1}{N_c}\right),\qquad C_R = g_s^2 (U^D_{d_R,1})^2 \frac{m_{\tilde{g}}}{m_{\tilde{g}}^2 - m_{\tilde{b}_1}^2} \left(1 - \frac{1}{N_c}\right) \,.
\end{equation}
The same result can be obtained in the large $m_{\tilde{g}}$ limit by integrating out the gluino in the Lagrangian, neglecting the kinetic term.

As QCD is parity invariant, the hadronic matrix elements in~\eqref{eq:omegabraket} are equal.  We estimate them using the
vacuum insertion approximation.  In this approximation, we insert the
vacuum between the operators in all possible ways, giving
\cite{Gaillard:1974hs,Donoghue:1992dd}:
\begin{equation}
  \label{eq:vac-ins-approx}
  \langle \bar{\tilde{B}} | (\tilde{b}^{*}_i d^i) (\tilde{b}^{*}_j d^j) |\tilde{B} \rangle
  \approx 2 \left[\langle  \bar{\tilde{B}} | (\tilde{b}^{*}_i d^i) |0\rangle
  \langle 0| (\tilde{b}^{*}_j d^j) |\tilde{B} \rangle + \langle \bar{\tilde{B}} | (\tilde{b}^{*}_i d^j) |0\rangle
  \langle 0| (\tilde{b}^{*}_j d^i) |\tilde{B} \rangle\right] \,,
\end{equation}
where we indicate color indices explicitly, and there are two ways to obtain each of the terms, yielding a prefactor of 2.
The contraction with the color-neutral external state kills the terms
with $i \neq j$ in the second term.  Exactly one in every $N_c$ terms has $i
= j$, so we get the relation:
\begin{equation}
  \label{eq:singlet-octet}
  \langle \bar{\tilde{B}} | (\tilde{b}^{*}_i d^j) |0\rangle  \langle 0|
  (\tilde{b}^{*}_j d^i) |\tilde{B} \rangle = \frac{1}{N_c} \langle \bar{\tilde{B}} | (\tilde{b}^{*}_i d^i) |0\rangle  \langle 0|
  (\tilde{b}^{*}_j d^j) |\tilde{B} \rangle \,.
\end{equation}
Our result is thus:
\begin{eqnarray}
  \label{eq:vac-ins-final}
  \langle \bar{\tilde{B}}| (\tilde{b}^* d) (\tilde{b}^* d) |\tilde{B} \rangle & = &
  \langle \bar{\tilde{B}} | (\tilde{b}^* d^{c \dag}) (\tilde{b}^* d^{c \dag}) |\tilde{B} \rangle \nonumber\\
  ~ & \approx & 2 \frac{N_c + 1}{N_c} \langle \bar{\tilde{B}} | (\tilde{b}^{*}_i d^i) |0\rangle
  \langle 0| (\tilde{b}^{*}_j d^j) |\tilde{B} \rangle \equiv 2 \frac{N_c +
  1}{N_c} f_{\tilde{B}}^2 m_{\tilde{B}} \,.
\end{eqnarray}

The mesino decay constant $f_{\tilde{B}}$ can be estimated using the $B$ meson decay constant
and assuming heavy quark symmetry.  Up to
threshold corrections, the relationship is given by \cite{Manohar:2000dt}:
\begin{equation}
  \label{eq:heavy-squark-quark}
  f_{\tilde{B}} = f_B
  \sqrt{\frac{m_b}{m_{\tilde{b}_1}}}
  \left(\frac{\alpha_s(m_b)}{\alpha_s(m_t)}\right)^{6/23} \left(\frac{\alpha_s(m_t)}{\alpha_s(m_{\tilde{b}_1})}\right)^{6/21}.
\end{equation}
Using the latest values of $f_B = 190.6~{\rm MeV}$
\cite{Laiho:2009eu,lat-averages} and $\alpha_s(m_Z) = 0.1184$
\cite{pdg}, the $\overline{MS}$ quark masses $m_b =
4.19~{\rm GeV}$ and $m_t = 160~{\rm GeV}$ \cite{pdg}, as well as
with a numerical solution to the NNNLO beta function for $\alpha_s$
\cite{Chetyrkin:1997sg}, which we evaluate at
$m_{\tilde{b}_1}=300~{\rm GeV}$, we find a value of $f_{\tilde{B}} =
28.7~{\rm MeV}$.

Putting these pieces together, we arrive at our final expression:
\begin{equation}
  \label{eq:deltam-mesino}
  \Delta m = g_s^2 \left|(U^{D}_{d_L,1})^2 + (U^{D}_{d_R,1})^2\right|
  f_{\tilde{B}}^2(m_{\tilde{b}_1}) \left(1 - \frac{1}{N_c^2}\right)
  \frac{m_{\tilde{g}}}{m_{\tilde{g}}^2 - m_{\tilde{b}_1}^2}\,,
\end{equation}
with $f_{\tilde{B}}$ given by \eqref{eq:heavy-squark-quark}. This
agrees with~\cite{Sarid:1999zx} up to a factor of 8 and the dependence
on the CP-violating phase in the squark mixing matrix.  This result
has some hadronic uncertainty, which we estimate to be of order
$10~\%$ based on estimates of the validity of the same approximations
for the $B$ meson systems.

 \end{appendix}

\end{document}